# Nuclear Magnetic Resonance and Magnetization Studies of the Ferromagnetic Ordering Temperature Suppression in Ru Deficient SrRuO$_3$

Z. H. Han,[a] J. I. Budnick,[a] M. Daniel,[a] W. A. Hines,[a,*] D. M. Pease,[a] P. W. Klamut,[b,c] B. Dabrowski,[b,c] S. M. Mini,[b,c] M. Maxwell,[b] C. W. Kimball[b]

[a]*Department of Physics and Institute of Materials Science, University of Connecticut, Storrs, CT 06269, USA*

[b]*Department of Physics, Northern Illinois University, DeKalb, IL 60115, USA*

[c]*Materials Science Division, Argonne National Laboratory, Argonne IL 60439, USA*

**Abstract**

The synthesis of SrRuO$_3$ under high-pressure oxygen produces a nonstoichiometric form with randomly distributed vacancies on the Ru-sites, along with a significantly reduced ferromagnetic ordering temperature. In order to gain additional insight into the suppression of the ferromagnetism, local studies utilizing $^{99,101}$Ru zero-field spin-echo NMR, and Ru $K$-edge XAFS, along with complimentary magnetization and x-ray diffraction measurements, have been carried out on samples of SrRuO$_3$ annealed at both ('ambient') atmospheric pressure and 'high-pressure' oxygen (600 atm). Consistent with previous work, the NMR spectrum for 'ambient' SrRuO$_3$ consists of two well-defined peaks at 64.4 MHz and 72.2 MHz corresponding to the $^{99}$Ru and $^{101}$Ru isotopes, respectively, and a hyperfine field of 329 kG. Although the magnetization measurements show a lower ferromagnetic ordering temperature for the 'high-pressure' oxygen sample (90 K compared to 160 K for the 'ambient' sample), the NMR spectrum shows no significant shift in the two peak frequencies. However, the two peaks exhibit considerable broadening, along with structure on both the low and high frequency sides which is believed to be quadrupolar in origin. Analysis of the Ru $K$-edge XAFS reveals more disorder in the Ru-O bond for the 'high-pressure' oxygen sample compared to the 'ambient' sample. Furthermore, XANES of Ru $K$-edge analysis indicates no difference in the valence of Ru between the two samples. The magnetic behavior indicates the existence of some vacancies on the Ru sites for the 'high-pressure' oxygen sample.

* Corresponding author. Tel.: +1-860-486-2343; Fax: +1-860-486-3346; e-mail: hines@uconnvm.uconn.edu

# 1. INTRODUCTION

Recently, there has been considerable research interest in ruthenate systems with the $ABO_3$ perovskite and related crystal structures. These 4d transition-metal oxides are found to have very interesting physical properties such as superconductivity, ferromagnetic ordering, and large magnetoresistance. Furthermore, these physical properties depend sensitively on the composition and synthesis conditions. Of crucial importance is an understanding of the electronic, magnetic, and atomic structures, and their relationship to the physical properties.

Among the ruthenate oxides, strontium ruthenate, $SrRuO_3$, which has a distorted perovskite structure, is a rare example of 4d transition-metal oxide which is ferromagnetic. Since its crystal structure is compatible with both high-$T_c$ superconductors and ferroelectric compounds, it has become a candidate for technological applications [1]. $SrRuO_3$ exhibits metallic-like conductivity with a behavior that is characterized as a 'bad metal' [2]. It is known to be a highly correlated, narrow-band metallic ferromagnet with an ordering temperature near 160 K. Due to the large crystal-field splitting of the energy levels, which typically occurs with 4d transition metals, the $Ru^{4+}$ ion is in the low-spin $S = 1$ state. Values have been reported for the ordered Ru magnetic moment in the ferromagnetic state which range from 0.8 to 1.6 $m_B$, although the magnetization has not reached saturation for the highest fields utilized [3]. This lack of saturation is consistent with itinerant ferromagnetism [2]. Furthermore, measurements on single-crystal specimens of $SrRuO_3$ dramatically illustrated an anisotropy in terms of the in-plane versus out-of-plane magnetization, with the [001] direction being the hard axis [3]. On the other hand, the paramagnetic state is characterized by Curie-Weiss behavior with a moment value of approximately 2.8 $m_B$ which is consistent with $S = 1$.

The substitution of isovalent Ca for Sr in $SrRuO_3$ maintains the metallic-like conductivity and perovskite structure while suppressing the magnetic ordering temperature $T_c$ [4]. This suppression has been attributed to an increased orthorhombic distortion and a larger deviation of the Ru-O-Ru bond angle from 180° [5]. Also, the substitution of 12 at.% Na for Sr increases the Ru formal valance to +4.12 and dramatically suppresses $T_c$ [6]. Somewhat reduced ordering temperatures for thin films of $SrRuO_3$ deposited on substrates have occurred. This has been attributed to strain effects arising from mismatched lattice parameters [7]. In addition, a low-$T_c$ magnetically ordered phase of $SrRuO_3$ has been reported which apparently depends on the synthesis temperature [8]. In an earlier work, it was reported that the synthesis of $SrRuO_3$ under high-pressure oxygen produces a nonstoichiometric form with randomly distributed vacancies on the Ru-sites, along with a significantly reduced ferromagnetic ordering temperature [9]. The decrease in the ordering temperature was attributed to an increase in the Ru formal valence along with structural disorder. In order to understand the microscopic origin of the suppression of the ferromagnetism in $SrRuO_3$, local studies utilizing $^{99,101}$Ru zero-field spin-echo nuclear magnetic resonance (NMR) and Ru $K$-edge x-ray absorption fine structure (XAFS), along with complimentary magnetization and x-ray diffraction, have been carried out and are reported here.

# 2. EXPERIMENTAL APPARATUS AND PROCEDURE

Polycrystalline $SrRuO_3$ was synthesized from a stoichiometric mixture of $SrCO_3$ and $RuO_2$ using a solid state reaction method. The material was reground and fired in air several times at various temperatures up to 1100 °C followed by slow cooling to 400 °C. A typical 2 gram-size batch of the parent $SrRuO_3$ material was subdivided and processed under the following conditions: (1) annealed twice under flowing oxygen at 1050 °C for 10 h ('ambient') and (2) annealed twice under 600 atm oxygen at 1050 °C for 10 h ('high-pressure'). In both cases, the material was ground to a course powder in preparation for the NMR and XAFS experiments. X-ray diffraction (XRD) analysis, which was carried out using a Bruker powder diffractometer and Cu $K\alpha$ radiation ($\lambda = 1.5418$ Å), confirmed that the powders were single phase. In addition, magnetization measurements were carried out on a Quantum Design MPMS

SQUID magnetometer for temperatures 5 K ≤ *T* ≤ 340 K and magnetic fields -50 kG ≤ *B* ≤ +50 kG. The effective oxygen content was studied using thermogravimetric measurements on a Cahn TG171 thermobalance.

In this work, detailed zero-field (no applied field) spin-echo [99,101]Ru NMR spectra were obtained at 1.3 K and 4.2 K over the frequency range from 45 to 105 MHz using a modified Matec model 5100 mainframe and model 525 gated RF amplifier in combination with a model 625 broadband receiver with phase sensitive detection. The NMR signals were optimized by adjusting the excitation power in a 2.0 $\mu s$ – 20 $\mu s$ - 2.0 $\mu s$ spin-echo pulse sequence having a repetition rate of 20 Hz. Usable spectra were obtained by averaging the NMR signals 500 to 1,000 times at 0.25 MHz intervals across the spectrum. In addition, measurements of the spin-spin relaxation time $T_2$ were made at the (two) peak frequencies by varying the separation *t* between the two RF pulses. Temperatures in the liquid helium range were obtained by pumping with a conventional double dewar system. The reader is referred to Zhang et al. [10] and references therein for details concerning the pulsed NMR apparatus, experimental procedure, and data acquisition and analysis.

The XAFS experiments were carried out in the transmission mode at the X-11A beamline of the National Synchrotron Light Source, Brookhaven National Laboratory, using a double-crystal Si (111) monochromator. The apparatus is described in detail elsewhere [11]. For the two samples, the powder was rubbed onto Kapton tape which was used to create a stack of layers in order to get an appropriate edge jump and to avoid systematic errors for the thickness effect. In this work, measurements of the Ru *K*-edge were carried out. Data reduction consisted of pre-edge and post-edge background subtraction, step normalization at the edge, extraction of the XAFS $\chi(k)$ function, and subsequent Fourier transform (FT) from *k* to *r* space. It was performed using the UWAFS 3.0 data analysis package [12].

X-ray absorption near edge spectroscopy (XANES) measurements of the Ru *K*-edge (22.117 keV) were performed in transmission mode at 12BM BESSRC beamline at the Advanced Photon Source, Argonne National Laboratory, using a double-crystal Si (111) monochromator. For comparison purposes, a standard ($RuO_2$) was used. Samples and the powder standard were prepared as described in the preceding paragraph.

## 3. RESULTS AND ANALYSIS

The magnetization results, which are summarized in Table 1, include parameters from the ferromagnetic state below the ordering temperature as well as the paramagnetic state above the ordering temperature. For both the 'ambient' and 'high-pressure' samples (column one), a direct measure of the magnetic ordering temperature $T_c$ was made by observing the zero-field-cooled and field-cooled dc magnetization in a field of 50 G (see Fig. 1). The values of $T_c$ = 160 K and 90 K (column two) obtained for the 'ambient' and 'high pressure' samples, respectively, are consistent with results obtained previously from ac susceptibility measurements (see the insert in Fig. 1; also, the values are listed in column three of Table 1) [9]. For comparison, the low temperature magnetic moment per Ru atom in the ferromagnetically-ordered state measured at 5.0 K and 50 kG (see Fig. 2) is listed in column four. These values assume complete occupancy of the Ru sites. The measurements of the complete hysteresis loops for both the 'ambient' and 'high-pressure' samples indicated a larger value of the coercive field for the latter (column five). This suggests an enhanced pinning of the domain walls in the 'high-pressure' sample. Columns six and seven list the parameters obtained from a Curie-Weiss fit (see Fig. 3) to the magnetic susceptibility in the paramagnetic state using

$$\chi(T) = \frac{n\mu_{eff}^2}{3k_B(T-\Theta)} + \chi_0, \qquad \text{eq. (1)}$$

where *n* is the concentration of Ru moments, $\mu_{eff}$ is the effective Ru moment (magnitude), $k_B$ is the Boltzmann constant, $\Theta$ is the Curie-Weiss temperature, and $\chi_0$ is a temperature independent term which reflects the core diamagnetism, Landau diamagnetism, and Pauli paramagnetism. The susceptibility values used in the Curie-Weiss fits were obtained at various temperatures above the ordering temperature by measuring the magnetization as a function of magnetic field in order to insure completely linear behavior through the origin. For both the 'ambient' and 'high-pressure' samples, the Curie-Weiss behavior was observed over a 125 K range starting 55 K above the ordering temperature (160 K and 90 K, respectively). The calculated magnetic moment values of $2.6_6$ $m_B$ and $2.6_1$ $m_B$ for the 'ambient' and 'high-pressure' samples, respectively, assume complete occupancy of the Ru sites. Although the uncertainty involved in the Currie-Weiss fit is comparable to the difference in the moment values, the lower value for the 'high pressure' oxygen sample is consistent with some vacancies at the Ru sites. Finally, column eight lists coefficients which were obtained by fitting the temperature dependence of the magnetization measured at 10 kG to the Bloch $T^{3/2}$ law, i.e.,

$$M(T)/M(0) = 1 - AT^{3/2}. \qquad \text{eq. (2)}$$

For comparison, $A_{90}/A_{160} = 1.9$ while $(160 \text{ K} / 90 \text{ K})^{3/2} = 2.4$.

Figure 4 shows the zero-field spin-echo NMR spectrum obtained for the 'ambient' sample at 4.2 K and 'high-pressure' sample at 1.3 K, respectively. Consistent with earlier work, the NMR spectrum for 'ambient' $SrRuO_3$ consists of two well-defined peaks at 64.4 MHz and 72.2 MHz corresponding to the $^{99}$Ru and $^{101}$Ru isotopes, respectively, and a hyperfine field of 329 kG [13]. For the $^{99}$Ru isotope, $g = 0.19645$ MHz/kG, $I = 5/2$, and 12.7% abundance, while for the $^{101}$Ru isotope, $g = 0.22018$ MHz/kG, $I = 5/2$, and 17.1% abundance [14]. For the 'high-pressure' sample, although the magnetization measurements show a lower ferromagnetic ordering temperature (90 K compared to 160 K for the 'ambient' sample), the NMR spectrum still shows the two peaks at 64.4 MHz and 72.2 MHz which are attributed to $^{99}$Ru and $^{101}$Ru, respectively, i.e., there is no significant shift. However, the two peaks exhibit considerable broadening, along with structure on both the low and high frequency sides which is believed to be quadrupolar in origin. These peak frequencies and the corresponding hyperfine field value of 329 kG are consistent with the $Ru^{4+}$ or low-spin ($S = 1$) valence state. The $^{99,101}$Ru NMR spectrum is qualitatively similar to those observed for the Ru sites having the $Ru^{4+}$ valence state in both $RuSr_2GdCu_2O_8$ [15] and $RuSr_2YCu_2O_8$ [16]. For these two systems, NMR spectra have been obtained for both $Ru^{4+}$ and $Ru^{5+}$. The spectra are characterized by a large internal hyperfine field perturbed by a smaller quadrupole interaction in which the Ru sites are located at a uniaxially symmetric position in distorted $RuO_6$ octahedra. The observed quadrupole interactions ($n_Q$) are large enough to account for the broadening shown in Fig. 4 for the 'high-pressure' sample. Furthermore, the effect of the high-pressure oxygen treatment is much more severe on the $^{101}$Ru peak, which is consistent with the fact that the quadrupole moment for $^{101}$Ru is almost six times larger than that for $^{99}$Ru. A search was made over the frequency range 100 MHz to 150 MHz in order to look for the existence of $Ru^{5+}$ or high-spin ($S = 2$) valence state. Unlike the results reported for both $RuSr_2GdCu_2O_8$ [15] and $RuSr_2YCu_2O_8$ [16], there was no NMR evidence for the $Ru^{5+}$ valence state in the 'high-pressure' sample. Measurements of the (homogeneous) spin-spin relaxation time $T_2$ were made at both peaks for the two samples. For the 'ambient' sample, the relaxation behavior could be described by a single exponential with $T_2 = 1,100$ $ms$ and 920 $ms$ for the $^{99}$Ru and $^{101}$Ru peaks, respectively. On the other hand, the relaxation behavior for the 'high-pressure' sample could not be described by a single exponential; both peaks having fast and slow components. In this case, $T_2 = 200$ $ms$ (320 $ms$) and 170 $ms$ (350 $ms$) for the $^{99}$Ru and $^{101}$Ru peaks, respectively. Finally, it should be noted that NMR spin-echo signal was considerably weaker for the 'high-pressure' sample compared to that for the 'ambient' sample, which required operation at 1.3 K. Of course, the severe broadening accounts for a large part of the signal reduction; however, the NMR enhancement factor appeared to be reduced as well for the 'high-pressure' sample. This is consistent with the increased coercive field observed for the 'high-pressure' sample and suggests an enhanced pinning of the domain walls.

The polycrystalline 'ambient' and 'high-pressure' $SrRuO_3$ samples were characterized by Cu $Ka_I$ XRD. Both samples showed the $GdFeO_3$-like orthorhombic structure. Furthermore, the values obtained for the lattice parameters $a$, $b$, and $c$ were the same for the two samples within the accuracy of the measurement (see Fig. 5). The values were also consistent with those reported previously for SrRuO$_3$ [17]. In an attempt to observed any small subtle differences between the XRD patterns for the 'high-pressure' and 'ambient' samples, a careful scan was made of two high-angle lines: (1) near 67° which consists of the (400) and (224) reflections and (2) near 77° which consists of the (116) reflection. Within the accuracy of the measurement, the scans for the 'ambient' and 'high-pressure' samples are identical (see the insert in Fig. 5).

Finally, analysis of the Ru *K*-edge XAFS indicates more disorder in the Ru-O bond for the 'high-pressure' sample compared to the 'ambient' sample. A close inspection of the Ru *K*-edge XANES for each sample showed no difference in the shape of the spectra nor shift in energy between the 'ambient' and 'high-pressure' samples. It may be inferred from the lack of energy shift that the overall Ru valence is the same in each sample. The near identical shape of the edge spectrum for each sample is indicative of parity in Ru molecular position and coordination. This is consistent the XRD observations.

Work at University of Connecticut was supported by the NSF (DMR-9705136), and at Northern Illinois University by the NSF (DMR-0105398) and by the State of Illinois under HECA. SMM acknowledges support by NSF (CHE-9871246). Use of the Advanced Photon Source BESSRC-CAT 12-BM beamline was supported by the U.S. Department of Energy, Basic Energy Sciences-Materials Science under contract No. W-31-109-ENG-38.

Table I

Magnetic Properties of SrRuO$_3$

| Sample | $T_c^{dc}$ (K) | $T_c^{ac}$ (K) | $m_r$ ($m_B$) | $H_c$ (Oe) | $\Theta$ (K) | $m_{\text{eff}}$ ($m_B$) | $A$ (K$^{-3/2}$) |
|---|---|---|---|---|---|---|---|
| 'ambient' | 160 | 161 | 1.25 | 3,500 | 153 | 2.6$_6$ | 2.7×10$^{-4}$ |
| 'high-pressure' | 90 | 86 | 0.86 | 4,500 | 98 | 2.6$_1$ | 5.1×10$^{-4}$ |

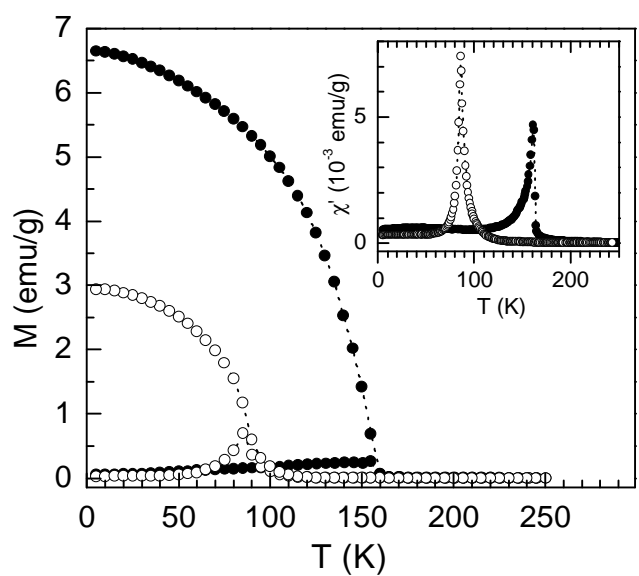

Fig. 1.

Zero-field-cooled and field-cooled dc magnetization in a field of 50 G versus temperature for the SrRuO$_3$ samples: solid circles - 'ambient'; open circles - 'high-pressure'. The ferromagnetic ordering temperature in the 'high-pressure' sample (90 K) is suppressed with respect to the 'ambient' sample (160 K). The insert shows results obtained previously from ac susceptibility measurements [9].

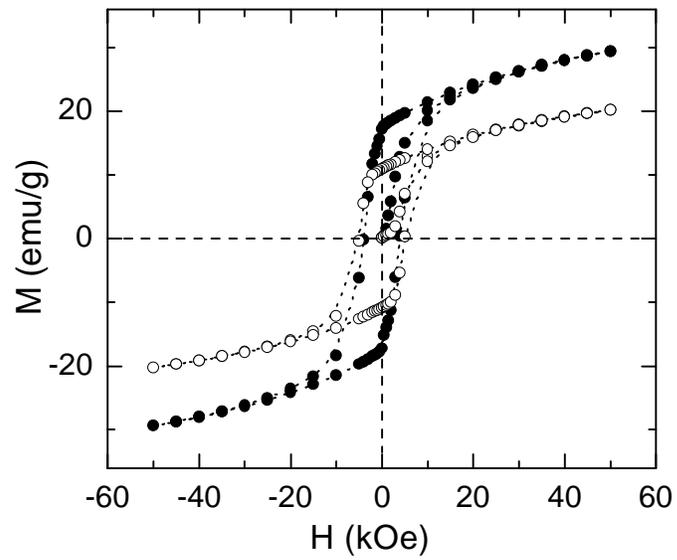

Fig. 2.

Full hysteresis loops obtained at 5.0 K for the SrRuO$_3$ samples: closed circles - 'ambient'; open circles - 'high-pressure'.

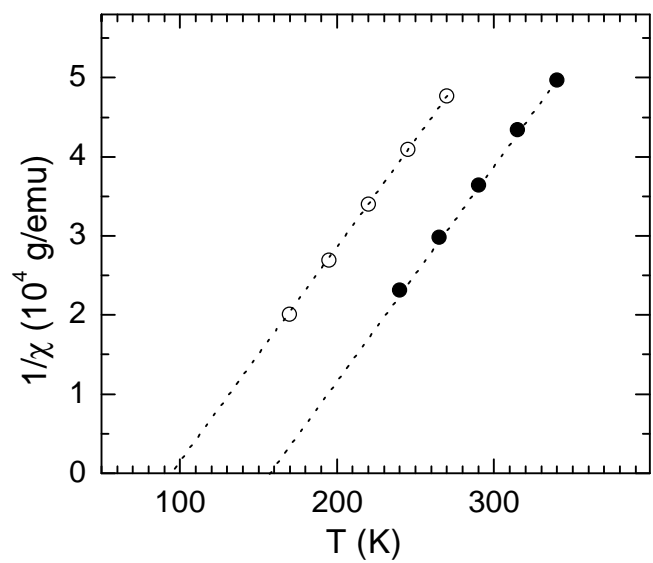

Fig. 3.

Fits to the Curie-Weiss law for the $SrRuO_3$ samples: closed circles - 'ambient'; open circles - 'high-pressure'.

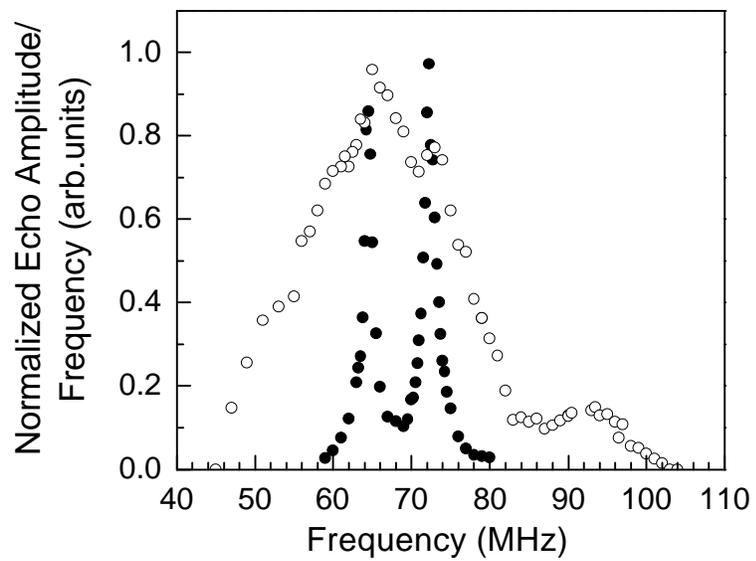

Fig. 4.

Zero-field spin-echo NMR spectra for $SrRuO_3$: closed circles - obtained at 4.2 K for the 'ambient' sample; open circles - obtained at 1.3 K for the 'high-pressure' sample. The peaks at 64.4 MHz and 72.2 MHz are attributed to $^{99}Ru$ and $^{101}Ru$, respectively, with a hyperfine field of 329 kG.

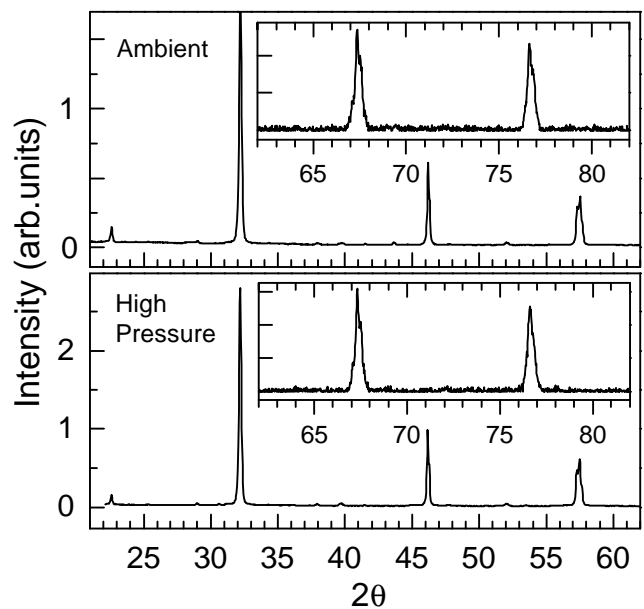

Fig. 5.

X-ray diffraction scans for the SrRuO$_3$ samples: upper - 'ambient'; lower – 'high-pressure'. Additional scans were made for the (400), (224), and (116) high-angle peaks (insert). Both samples showed the GdFeO$_3$-like orthorhombic structure. Within the accuracy of the measurement, the XRD patterns for the two samples are identical.